\documentclass[a4paper,12pt]{article}

\usepackage{amsmath}
\usepackage[psamsfonts]{amssymb}
\usepackage{rsfs}
\usepackage{bm}

\usepackage{cite}
\usepackage[dvips]{graphicx}
\usepackage{color}

\makeatletter
\@addtoreset{equation}{section}
\makeatother

\begin{document} 

\begin{titlepage}

\baselineskip 10pt
\hrule 
\vskip 5pt
\leftline{}
\leftline{Chiba Univ./KEK Preprint
          \hfill   \small \hbox{\bf CHIBA-EP-170}}
\leftline{\hfill   \small \hbox{\bf KEK Preprint 2007-73}}
\leftline{\hfill   \small \hbox{January 2008}}
\vskip 5pt
\baselineskip 14pt
\hrule 
\vskip 1.0cm
%
\vskip 0.3cm
\centerline{\Large\bf  
Proving  Abelian dominance  
}
\vskip 0.3cm
\centerline{\Large\bf  
in the Wilson loop operator 
}
\vskip 0.3cm

\vskip 1.0cm

\centerline{{\bf 
Kei-Ichi Kondo $^{\dagger,{1}}$  
and
Akihiro Shibata$^{\flat,{2}}$  
}}  
\vskip 0.5cm
\centerline{\it
${}^{\dagger}$Department of Physics, Graduate School of Science, 
}
\centerline{\it
Chiba University, Chiba 263-8522, Japan
}
\vskip 0.3cm
\centerline{\it
${}^{\flat}$Computing Research Center, High Energy Accelerator Research Organization (KEK)   
}
\vskip 0.1cm
\centerline{\it
\& 
Graduate Univ. for Advanced Studies (Sokendai), 
Tsukuba 
305-0801, 
Japan
}
\vskip 1cm

\begin{abstract}
 We give a gauge-independent definition of Abelian dominance in the Wilson loop operator and a constructive proof of the Abelian dominance through a non-Abelian Stokes theorem via  lattice regularization. 
We obtain a necessary and sufficient condition for the Abelian dominance in the Wilson loop operator in the fundamental representation. 
In the continuum limit, the gauge field is decomposed such that the Abelian dominance is given as an exact operator relation,  leading to the exact (100\%) Abelian dominance.  
On a lattice, we estimate the deviation from the exact Abelian dominance due to non-zero lattice spacing. 
In order to obtain the best Abelian dominance on a lattice by minimizing the deviation, we discuss how to decompose  the gauge field variable into the dominant part and the remaining one to be decoupled on a lattice. 

\end{abstract}

Key words:   Wilson loop, Abelian dominance,  Stokes theorem, quark confinement, Yang-Mills theory, 
maximal Abelian gauge,

\vskip 0.5cm

PACS: 12.38.Aw, 12.38.Lg 
\hrule  
\vskip 0.1cm
${}^1$ 
  E-mail:  {\tt kondok@faculty.chiba-u.jp}
  
${}^2$ 
  E-mail:  {\tt akihiro.shibata@kek.jp}

\par 
\par\noindent


\vskip 0.5cm

\newpage
\pagenumbering{roman}
\tableofcontents




\end{titlepage}


\pagenumbering{arabic}

\baselineskip 14pt


\section{Introduction}
 
Quark confinement is a fundamental problem to be solved in theoretical physics. However, we have not yet obtained a satisfactory derivation of quark confinement.  Even if a mathematical proof of quark confinement was given, we wish to know the fundamental  mechanism behind the phenomenon of quark confinement from the physical point of view.  
In this paper we consider quark confinement based on the Wilson criterion \cite{Wilson74} in Yang-Mills theory \cite{YM54}: quark confinement is realized by the linear potential between a pair of a quark and an antiquark, if the Wilson loop average exhibits the area decay:
\begin{align}
 W(C) := \Big\langle 
W_C[\mathscr{A}]
\Big\rangle_{\rm YM} \sim e^{-\sigma_{NA} |S|}
 ,
\end{align}
where $W_C[\mathscr{A}]$ is the Wilson loop operator defined by 
\begin{align}
 W_C[\mathscr{A}] :=  
{\rm tr} \left[ \mathscr{P} \exp \left\{ ig \oint_{C} dx^\mu \mathscr{A}_\mu(x) \right\} \right]
 .
\end{align}

In view of this, the most promising mechanism for quark confinement is believed to be  the dual superconductivity, which is an idea proposed long ago \cite{dualsuper}.  
However, it was not straightforward to realize this idea in Yang-Mills theory to demonstrate the dual superconductivity. 
It was 't Hooft \cite{tHooft81} who proposed the procedure called the \textbf{Abelian projection} for substantiating this idea of the dual superconductivity in the Yang-Mills theory:
The Abelian projection enables one to realize or extract magnetic monopole even in the pure Yang-Mills theory which does not involve the (adjoint) scalar field, in sharp contrast to the 'tHooft-Polyakov magnetic monopole in the Georgi-Glashow model.  
The original Abelian projection is regarded as a partial gauge fixing, which breaks the original gauge group $G=SU(N)$ to the maximal torus subgroup $H=U(1)^{N-1}$.  
Although there exist many ways to perform the partial gauge fixing $G \rightarrow H$,  recent investigations have shown that quark confinement based on the dual superconductor picture can be effectively realized in  a specific gauge called the \textbf{maximal Abelian gauge (MAG)} \cite{KLSW87}.
For instance, 
 the MAG in the differential form in the continuum formulation has the form:
$
  [\partial_\mu \delta^{ab} - g \epsilon^{ab3}A_\mu^3(x)] A_\mu^b(x) = 0 \ (a,b=1,2) 
   ,
$
for the decomposition of the SU(2) gauge field $\mathscr{A}_\mu(x)$ into the diagonal component $A_\mu^3(x)$ and off-diagonal ones $A_\mu^a(x)$  
distinguished by the Pauli matrices $\sigma_A$ ($A=1,2,3$):
\begin{equation}
\mathscr{A}_\mu(x)
= A_\mu^3(x)  \sigma^3/2 + A_\mu^a(x)  \sigma^a/2  \ (a=1,2) 
 .
 \label{SU2-decomp}
\end{equation}
The non-perturbative way of imposing the MAG is to minimize the functional 
$\int d^4x \frac12 A_\mu^a A_\mu^a$ with respect to the gauge transformation, just as the Lorenz-Landau gauge fixing is obtained from minimizing the functional
$\int d^4x \frac12 \mathscr{A}_\mu^A \mathscr{A}_\mu^A$.

Numerical simulations on a lattice  have discovered \cite{SY90} that the Abelian-projected Wilson loop $W_{\rm Abel}(C) $ defined by the diagonal component $A_\mu^3$ alone exhibits the area decay law, if the MAG is imposed on the SU(2) Yang-Mills theory on a lattice:%
\footnote{
If no gauge fixing were imposed, the Yang-Mills average of the non gauge-invariant operator would yield a trivial result (due to the Elitzur theorem).
}
\begin{align}
  W_{\rm Abel}(C) = \Big\langle \exp \left\{ i g \oint_{C} dx^\mu A_\mu^3(x) \right\} \Big\rangle_{\rm YM}^{\rm MAG} \sim e^{-\sigma_{Abel} |S|} 
   .
\end{align}
The most remarkable characteristics obtained in the MAG are dominances in the string tension: In numerical simulations, the string tension $\sigma_{Abel}$ obtained from the Abelian Wilson loop average $W_{\rm Abel}(C)$ reproduces almost all the value (95$\%$) of the string tension $\sigma_{NA}$ obtained from the original Wilson loop average $W(C)$ without gauge fixing, which is called the Abelian dominance in the string tension \cite{tHooft81,EI82}: 
\par
\textbf{Abelian dominance}\cite{SY90} $\Leftrightarrow$ $\sigma_{NA}  \sim \sigma_{Abel}$

Moreover, if the diagonal gauge potential is decomposed into the so-called  photon and monopole parts: 
$
  A_\mu^3 = \text{Photon~part}+\text{Monopole part} 
$, 
such that only the monopole part gives the non-vanishing magnetic charge, 
then the monopole part reproduces almost all the value (95$\%$) of the string tension $\sigma_{Abel}$, which is called the monopole dominance in the string tension: 
\par
\textbf{Monopole dominance}\cite{SNW94} $\Leftrightarrow$ $\sigma_{Abel}  \sim \sigma_{monopole}$

Therefore,  MAG is believed to be the most efficient way to perform the Abelian projection to demonstrate the dual superconductivity. 
These results are indeed remarkable progress towards the goal of understanding quark confinement based on the dual superconductivity in Yang-Mills theory. 
See \cite{CP97} for a review.

However, we are still unsatisfactory with the following reasons.
\begin{itemize}
\item
The Abelian projection and MAG break  SU(2) color symmetry  explicitly. 

This is inconvenient, because we consider that quark confinement should be understood as a special case of color confinement and the color confinement is well defined only if the color symmetry is preserved.  If the color symmetry is broken, we lose a chance of explaining color confinement as an extension of quark confinement. 

\item
The Abelian dominance has never been observed (at least so far) in  gauge fixings other than MAG in numerical simulations. 

This causes the dubious impression that the dual superconductivity obtained in MAG might be a gauge artifact. 

\end{itemize}

In order to establish the dual superconductivity  as a gauge-invariant concept, we should cure these shortcomings associated with the MAG.
In order to reconsider the meaning of the Abelian projection and the MAG with the resulting Abelian dominance, we consider the Wilson loop operator $W_C[\mathscr{A}]$ for the SU(N) gauge connection $\mathscr{A}$ 
whose vacuum expectation value is the Wilson loop average $W(C)= \langle W_C[\mathscr{A}]  \rangle_{\rm YM}$.

We define a \textit{gauge-independent Abelian dominance in the SU(2) Wilson loop} as follows.  

(i) Consider the decomposition of the original SU(2) gauge field into two parts:
\begin{equation}
 \mathscr{A}_\mu(x) = \mathscr{V}_\mu(x) + \mathscr{X}_\mu(x)  
  ,
  \label{decomp-0}
\end{equation} 
where $\mathscr{V}_\mu(x)$ transforms just the same way as $\mathscr{A}_\mu(x)$ under the gauge transformation.  
Then one of the variables $\mathscr{X}_\mu(x)$ 
does not contribute to the Wilson loop operator at all, i.e., 
\begin{equation}
 W_C[\mathscr{A}]= {\rm const.}  W_C[\mathscr{V}]
 .
\end{equation} 

(ii) The non-Abelian field strength $\mathscr{F}_{\mu\nu}[\mathscr{V}](x)$ of the  restricted field $\mathscr{V}_\mu(x)$ has the form  $\mathscr{F}_{\mu\nu}[\mathscr{V}](x)=F_{\mu\nu}(x)\bm{n}(x)$ where an isovector $\bm{n}(x)$ specifies the Abelian direction and the magnitude $F_{\mu\nu}(x)$ is SU(2) gauge invariant.   
Once the Wilson loop operator is rewritten in terms of the surface integral over the surface bounding the closed loop $C$,  $W_C[\mathscr{V}]$ is written in terms of only the SU(2) invariant Abelian field strength $F_{\mu\nu}(x)$, i.e., $W_S[F]$: 
\begin{equation}
W_C[\mathscr{V}]=W_S[F]
    .
\end{equation}  
The general SU(N) case is given later for technical reasons.

For SU(2), the fact that the Cho-Faddeev-Niemi-Shabanov (CFNS) decomposition \cite{Cho80,FN98,Shabanov99} for (\ref{decomp-0}) leads to the Abelian dominance in this sense was already pointed out by Cho  \cite{Cho00} in the continuum formulation through the non-Abelian SU(2) Stokes theorem of the Diakonov-Petrov type \cite{DP96,FITZ00}. 
Therefore, the CFNS decomposition is a sufficient condition for the  Abelian dominance in the SU(2) Wilson loop operator.

In this paper, we consider a necessary and sufficient condition for the Abelian dominance  for  the SU(N) Wilson loop operator.  Then we give a constructive proof of the SU(N) version of the gauge-invariant Abelian dominance  for  the Wilson loop operator in the fundamental representation through the SU(N) non-Abelian Stokes theorem \cite{KondoIV,KT99,HU99,Kondo08}.  
Consequently, the strong or exact Abelian dominance holds for the Wilson loop operator itself with arbitrary shape and size, not restricted to the inter quark potential derived from the Wilson loop average. 
Since the  Abelian projection (\ref{SU2-decomp}) is obtained as a special gauge fixing from our gauge-invariant framework,  the Abelian dominance in the string tension as a coefficient of the linear part in the interquark potential immediately follows from the gauge-independent Abelian dominance just defined. 
 The relationship between the Wilson loop operator and magnetic monopole has been discussed in a separate paper \cite{Kondo08}.

This paper is organized as follows. 
In section 2, we start from  the lattice regularized version of the non-Abelian Stokes theorem and the lattice versions of the gauge field decomposition which have been constructed and developed  for SU(2) in \cite{KKMSSI05,IKKMSS06,SKKMSI07} and for SU(N) in \cite{KSSMKI08,SKKMSI07b} according to the continuum versions given in \cite{KMS06} and \cite{KSM08}, respectively. 
In sections 2 and 4, then, we obtain a necessary and sufficient condition for realizing the gauge-independent Abelian dominance in the SU(N) Wilson loop operator (in the fundamental representation).  Moreover, we show that this condition agrees with a set of the defining equations  which specify the SU(N) gauge field decomposition (\ref{decomp-0}): new versions of decomposition \cite{Kondo08,KSM08} for SU($N$) case ($N \ge 3$), including the CFNS decomposition for SU(2) case. 
Therefore, the strong or exact Abelian dominance in the Wilson loop is an immediate consequence of such decomposition (\ref{decomp-0}). 
In other words, the exact Abelian dominance gives a raison d'etre and a physical meaning of gauge covariant decomposition of the gauge field variable \textit{a l\'a}   CFNS. 
In section 3, we give some technical materials for constructing a non-Abelian Stokes theorem, which is extracted from \cite{Kondo08}.

It should be remarked that the exact Abelian dominance in the above sense is obtained only in the continuum limit of the lattice spacing going to zero $\epsilon \rightarrow 0$. Therefore, there exists some deviation on a lattice coming from non-zero lattice spacing $\epsilon >0$.  
In section 5, we give an estimation of this deviation up to  $O(\epsilon^2)$. 
The constructive approach enables us to compare the result of numerical simulations with the theoretical consideration, or even give a numerical proof of the non-Abelian Stokes theorem. 
Moreover this will shed new light on the role of the MAG for the Abelian dominance on a lattice as will be discussed in section 6. 
These are advantages of the constructive proof given in this paper.

\section{Lattice regularization and decomposition of lattice variable }

In this paper, we adopt a lattice approximation in which the $D$-dimensional continuum spacetime $\mathbb{R}^{D}$ is replaced by the $D$-dimensional Euclidean lattice $L_\epsilon=(\epsilon \mathbb{Z})^{D}$ with a lattice spacing $\epsilon$. 
Let $W_\gamma[U]$ be the parallel transporter along an oriented path $\gamma$ composed of oriented links on a lattice $L_\epsilon$.  
Then $W_\gamma[U]$ is obtained by the path-ordered product of the  gauge variable $U_\ell$ defined on each link $\ell \in \gamma$. 
Especially, the Wilson loop operator for a closed loop $C$ on a lattice is defined by the trace of $W_\gamma$ for a closed loop $\gamma=C$:
\begin{equation}
 W_C[U] = {\rm tr}(U_C)/{\rm tr}(\bm{1}) ,
 \quad 
 U_C := \mathscr{P} \prod_{\ell \in C}  U_\ell
  ,
\end{equation}
where $\mathscr{P}$ is the path-ordered symbol. 
If the lattice spacing $\epsilon$ is sufficiently small, the operator $W_C[U]$ on a lattice $L_\epsilon$ can be a good approximation for the  Wilson loop operator $W_C[\mathscr{A}]$ defined on the continuum spacetime $\mathbb{R}^{D}$.  

Now we consider decomposing the SU(N) gauge variable $U_\ell=U_{x,\mu}$ defined on an oriented link $\ell=<x,x+\mu> \in L_\epsilon$ into the product of two SU(N) variables $X_{x,\mu}$ and $V_{x,\mu}$ defined on the same lattice \cite{KSSMKI08}:
\begin{equation}
 U_{x,\mu} = X_{x,\mu} V_{x,\mu} \in SU(N) ,
 \quad X_{x,\mu}, V_{x,\mu} \in SU(N)
  .
  \label{decomp-1}
\end{equation}
In lattice gauge theories, the link variable $U_{x,\mu}$ obeys the well-known lattice gauge transformation:
\begin{equation}
  U_{x,\mu}  \rightarrow \Omega_{x} U_{x,\mu} \Omega_{x+\mu}^\dagger = U_{x,\mu}^\prime
  , \quad \Omega_{x} \in SU(N)
  \label{U-transf}
 .
\end{equation}
Here we require that $V_{x,\mu}$ is a new link variable which transforms like a usual gauge variable on a link $\ell=<x,x+\mu>$: 
\begin{equation}
  V_{x,\mu}  \rightarrow \Omega_{x} V_{x,\mu} \Omega_{x+\mu}^\dagger = V_{x,\mu}^\prime 
  , \quad \Omega_{x} \in SU(N)
   .
   \label{V-transf}
\end{equation}
For this gauge transformation to be consistent with the decomposition (\ref{decomp-1}), consequently, $X_{x,\mu}$ must behave like an adjoint matter field at the site $x$ under the gauge transformation:
\begin{equation}
  X_{x,\mu} (=  U_{x,\mu}V_{x,\mu}^\dagger )
 \rightarrow \Omega_{x} X_{x,\mu} \Omega_{x}^\dagger = X_{x,\mu}^\prime
  , \quad \Omega_{x} \in SU(N)
  .
  \label{X-transf1}
\end{equation}

The explicit form for the new lattice variables $X_{x,\mu}$ and $V_{x,\mu}$ in terms of the original link variable $U_{x,\mu}$ and the color field $\bm{m}_{x}$ has been given for SU(2) in \cite{IKKMSS06,SKKMSI07} and for SU(N) in \cite{KSSMKI08,SKKMSI07b}.

Now we  define the Wilson loop operator for the decomposed variable (\ref{decomp-1}).   We repeat the same steps as those taken in deriving a non-Abelian Stokes theorem of the Diakonov-Petrov type for the Wilson loop operator in the continuum \cite{Kondo08}. 
First, inserting the complete set of coherent states $| \xi_{x}, \Lambda \rangle$ at every site $x$ on the loop $C$: 
\begin{equation}
 {\bf 1} = \int \left| \xi_x, \Lambda \right> d\mu(\xi_x) \left< \xi_x, \Lambda \right|   ,
\end{equation}
and replacing the trace of an operator $\mathscr{O}$ with 
\begin{equation}
 \mathcal{N}^{-1} {\rm tr}(\mathscr{O}) 
 = \int d\mu(\xi_x) \left< \xi_x, \Lambda \right| \mathscr{O}\left| \xi_x, \Lambda \right> 
  ,
\end{equation}
we obtain
\begin{align}
W_C[U] =  
\prod_{x \in C} \int d\mu(\xi_x) 
\prod_{\ell=<x,x+\epsilon \hat\mu> \in C} \langle \xi_{x},\Lambda|U_{\ell} | \xi_{x+\epsilon \hat\mu}, \Lambda \rangle
 .
 \label{path-int-U}
\end{align}
For details on coherent states and a non-Abelian Stokes theorem, see \cite{Kondo08,KT99}. 
It should be remarked that the insertion was performed to give the matrix element in the form
$\langle \xi_{x},\Lambda|U_{\ell} | \xi_{x+\epsilon \hat\mu}, \Lambda \rangle$ 
for preserving the gauge invariance even after inserting the coherent state, instead of taking
$\langle \xi_{x+\epsilon \hat\mu},\Lambda|U_{\ell} | \xi_{x}, \Lambda \rangle$.
Applying the decomposition (\ref{decomp-1}), then, we have
\begin{align}
W_C[U] = W_C[XV] =
 \prod_{x \in C} \int d\mu(\xi_x) 
\prod_{\ell=<x,x+\epsilon \hat\mu> \in C} \langle \xi_{x},\Lambda| X_{x,\mu} V_{x,\mu} | \xi_{x+\epsilon \hat\mu}, \Lambda \rangle
\label{path-int-XV}
 .
\end{align} 
Next, we look for a set of requirements which guarantees that the full Wilson loop operator $W_C[U]$ defined originally in terms of the link variable $U_\ell$ agrees exactly with the restricted one $W_C[V]$ which is rewritten in terms of the new link variable $V_\ell$ up to a constant factor:
\begin{align}
W_C[U] = {\rm const.} W_C[V] .
\label{condition}
\end{align} 
Roughly speaking, this Abelian dominance follows if 
the variable $X_{x,\mu}$ does not contribute to the Wilson loop operator through the matrix element between the coherent states.

If a representation of the Wilson loop is chosen, a reference state $| \Lambda \rangle$ of the coherent state is fixed and the the stability group $\tilde{H}$ is determined \cite{Kondo08}. 
In order to realize the Abelian dominance (\ref{condition}), we impose two requirements:
\begin{enumerate}
\item[I)] For the stability group $\tilde{H}$, we require
\begin{subequations}
\begin{align}
   \xi_{x}^\dagger  V_{x,\mu} \xi_{x+\epsilon \hat\mu}  \in \tilde{H} 
  ,
  \label{req-I}
\end{align}
which implies by definition of the  stability group
\begin{align}
 \xi_{x}^\dagger  V_{x,\mu} \xi_{x+\epsilon \hat\mu} | \Lambda \rangle  =&   | \Lambda \rangle  e^{i\phi_{x,\mu}}
  ,
  \label{req-I1}
\end{align}
with
\begin{equation}
  e^{i\phi_{x,\mu}} = \langle  \Lambda|  \xi_{x}^\dagger  V_{x,\mu} \xi_{x+\epsilon \hat\mu} | \Lambda \rangle
 .
\end{equation}
\end{subequations}
\item[II)] For the matrix element of $X_{x,\mu}$ in the coherent state $| \xi_{x}, \Lambda \rangle$, we require
\begin{equation}
  \prod_{<x,x+\mu> \in C} \langle  \Lambda| \xi_{x}^\dagger  X_{x,\mu} \xi_{x} | \Lambda \rangle  
  \equiv   \prod_{<x,x+\mu> \in C} \langle \xi_{x}, \Lambda| X_{x,\mu} | \xi_{x}, \Lambda \rangle 
=:  \rho_C[X,\xi] = {\rm const.}
 ,
 \label{req-II}
\end{equation}
where the constant is a complex number which is independent of $X_{x,\mu}$ and $\xi_{x}$. 
\end{enumerate}

Now we show that the Abelian dominance (\ref{condition}) follows from the requirements (I) and (II) for the Wilson loop in any representation. Moreover, we show that \textit{a set of requirements (I) \& (II) is a necessary and sufficient condition for the Abelian dominance (\ref{condition}) for the Wilson loop in the fundamental representation}.  
We first observe that for any representation the matrix element in (\ref{path-int-U}) reads 
\begin{align}
 \langle \xi_{x},\Lambda|U_{\ell} | \xi_{x+\epsilon \hat\mu}, \Lambda \rangle
 = \langle \Lambda| \xi_{x}^\dagger U_{\ell}\xi_{x+\epsilon \hat\mu}  |  \Lambda \rangle
  ,
\end{align}
while the matrix element in (\ref{path-int-XV}) is cast into 
\begin{align}
\langle \xi_{x},\Lambda| X_{x,\mu} V_{x,\mu} | \xi_{x+\epsilon \hat\mu}, \Lambda \rangle
=   \langle  \Lambda| (\xi_{x}^\dagger  X_{x,\mu} \xi_{x}) (\xi_{x}^\dagger  V_{x,\mu} \xi_{x+\epsilon \hat\mu}) | \Lambda \rangle  
 ,
 \label{condition2}
\end{align}
where we have used%
\footnote{
For $g=\xi h \in G=SU(N)$, $h \in \tilde{H}$, $\xi \in G/\tilde{H}$, we have 
$\bm{1}=gg^\dagger=\xi h h^\dagger \xi^\dagger$.  For $G=SU(N)$, $\tilde{H}=U(N-1)$ in the minimal case and $\tilde{H}=U(1)^{N-1}$ in the maximal case. Therefore, we have $h h^\dagger=\bm{1}$ in both cases, which leads to $\xi  \xi^\dagger=\bm{1}$.  These exhaust all the cases for SU(2) and SU(3). We exclude the other intermediate cases which can occur for SU(N), $N \ge 4$. 
}
$\xi_{x}  \xi_{x}^\dagger=\bm{1}$.

For fundamental representations, in particular, it is shown  using (\ref{dd-element}) that the matrix element $\langle \xi_{x},\Lambda|U_{\ell} | \xi_{x+\epsilon \hat\mu}, \Lambda \rangle$ is equal to one of the diagonal elements of $\xi_{x}^\dagger U_{\ell}\xi_{x+\epsilon \hat\mu}$:
\begin{align}
 \langle \xi_{x},\Lambda|U_{\ell} | \xi_{x+\epsilon \hat\mu}, \Lambda \rangle
 = (\xi_{x}^\dagger U_{\ell}\xi_{x+\epsilon \hat\mu})_{dd} 
  ,
\end{align}
while the matrix element $\langle \xi_{x},\Lambda| X_{x,\mu} V_{x,\mu} | \xi_{x+\epsilon \hat\mu}, \Lambda \rangle$ is decomposed to the product:
\begin{align}
\langle \xi_{x},\Lambda| X_{x,\mu} V_{x,\mu} | \xi_{x+\epsilon \hat\mu}, \Lambda \rangle
=&   (\xi_{x}^\dagger  X_{x,\mu} \xi_{x}  \xi_{x}^\dagger  V_{x,\mu} \xi_{x+\epsilon \hat\mu})_{dd} 
\nonumber\\
=&   (\xi_{x}^\dagger  X_{x,\mu} \xi_{x})_{da} (\xi_{x}^\dagger  V_{x,\mu} \xi_{x+\epsilon \hat\mu})_{ad} 
 ,
 \label{XV-2}
\end{align}
where no sum is understood for the index $d$,  while the summation over $a$ ($a=1, \cdots, N$) should be understood. 
We find that the Wilson loop operator $W_C[U]$ is written using  the diagonal element $(\xi_{x}^\dagger U_{\ell}\xi_{x+\epsilon \hat\mu})_{dd}=\langle \xi_{x},\Lambda|U_{\ell} | \xi_{x+\epsilon \hat\mu}, \Lambda \rangle$ in (\ref{path-int-U}). 
Therefore, we wish to obtain the condition for the matrix element $\langle \xi_{x},\Lambda| X_{x,\mu} V_{x,\mu} | \xi_{x+\epsilon \hat\mu}, \Lambda \rangle$ to be written in terms of only  the diagonal element:
\begin{align}
  (\hat{V}_{x,\mu})_{dd}
 = (\xi_{x}^\dagger V_{\ell}\xi_{x+\epsilon \hat\mu})_{dd}
  = \langle \Lambda| \xi_{x}^\dagger V_{\ell}\xi_{x+\epsilon \hat\mu}  |  \Lambda \rangle
 = \langle \xi_{x},\Lambda|V_{\ell} | \xi_{x+\epsilon \hat\mu}, \Lambda \rangle
 .
\end{align}
In fact, only a diagonal element of $\hat{V}_{x,\mu}:=\xi_{x}^\dagger  V_{x,\mu} \xi_{x+\epsilon \hat\mu}$  contributes in (\ref{XV-2})  if and only if the requirement I (\ref{req-I1}) is satisfied. 
This is verified as follows.  
For a choice of the fundamental representation:
\begin{equation}
\left|  \Lambda \right> 
= (0, \cdots,0,1)^T
= \begin{pmatrix}
  0 \cr
  \vdots \cr
  0 \cr
  1  
 \end{pmatrix}
 ,
\end{equation}
the requirement I (\ref{req-I1}) is equivalent to the equality of the matrix element for arbitrary state $\Psi$ ($\left< \Psi \right| = (\Psi_{1}^*, \Psi_{2}^*, \cdots, \Psi_{N}^*) $):
\begin{align}
  \langle \Psi |\hat{V}_{x,\mu}|\Lambda \rangle 
= \sum_{a=1}^{N} \Psi_{a}^* (\hat{V}_{x,\mu})_{aN} 
=\Psi_{N}^*  e^{i\phi_{x,\mu}} 
  .
\end{align}
This is fulfilled if and only if 
\begin{align}
   (\hat{V}_{x,\mu})_{aN} = 0 \ (a=1, \cdots, N-1) ,
   \quad 
   (\hat{V}_{x,\mu})_{NN} = e^{i\phi_{x,\mu}} \ne 0
  .
\end{align}
By repeating this argument for $\hat{V}_{x,\mu}^\dagger$, we find that the elements in the last row and the last column are vanishing $(\hat{V}_{x,\mu})_{aN} = 0=(\hat{V}_{x,\mu})_{Na}$ ($a=1, \cdots, N-1$) except for a diagonal element $(\hat{V}_{x,\mu})_{NN}$ being non-vanishing, that is to say, the matrix $\hat{V}_{x,\mu}=\xi_{x}^\dagger  V_{x,\mu} \xi_{x+\epsilon \hat\mu}$ reduces to a block-diagonal matrix. 
Consequently, only the ($NN$) diagonal element of the matrix $\hat{X}_{x,\mu} :=\xi_{x}^\dagger  X_{x,\mu} \xi_{x}$ contributes to the Wilson loop operator. 
Under the requirement (I), therefore, we have
\begin{align}
\langle \xi_{x},\Lambda| X_{x,\mu} V_{x,\mu} | \xi_{x+\epsilon \hat\mu}, \Lambda \rangle
=&   (\xi_{x}^\dagger  X_{x,\mu} \xi_{x})_{NN} (\xi_{x}^\dagger  V_{x,\mu} \xi_{x+\epsilon \hat\mu})_{NN} 
\nonumber\\
=& \langle \xi_{x}, \Lambda| X_{x,\mu} | \xi_{x}, \Lambda \rangle \times 
\langle \xi_{x},\Lambda| V_{x,\mu} | \xi_{x+\epsilon \hat\mu}, \Lambda \rangle 
 .
\end{align}

Under the requirement (I), anyway, the full Wilson loop reads
\begin{align}
 W_C[U]  
=& \prod_{x \in C} \int d\mu(\xi_x) 
  \prod_{\ell=<x,x+\epsilon \hat\mu> \in C} \langle \xi_{x}, \Lambda| X_{x,\mu} | \xi_{x}, \Lambda \rangle
\nonumber\\& 
 \quad \quad \quad \quad \times
\prod_{\ell=<x,x+\epsilon \hat\mu> \in C} 
\langle \xi_{x},\Lambda| V_{x,\mu} | \xi_{x+\epsilon \hat\mu}, \Lambda \rangle 
  .
\end{align} 
The term
$\rho_C[X,\xi]:=\prod_{\ell=<x,x+\epsilon \hat\mu> \in C} \langle \xi_{x}, \Lambda| X_{x,\mu} | \xi_{x}, \Lambda \rangle$ 
depends on $\xi_{x}$ as well as $X_{x,\mu}$ on the loop $C$, and is responsible for the integration by the measure $d\mu(\xi_x)$. 
The requirement (II) allows us to factor out this term to obtain the Abelian dominance equality (\ref{condition}).
A refinement of the requirement (II) will be discussed later. 

Once  the new lattice variables $V_{x,\mu}$ and $X_{x,\mu}$ are identified with elements of the SU(N) Lie group, they can be related to the $su(N)$ Lie algebra $\mathscr{V}_\mu(x)$ and $\mathscr{X}_\mu(x)$ respectively:%
\footnote{ 
Here we have assigned the Lie algebra $\mathscr{A}_\mu$ and $\mathscr{V}_\mu$ to the Lie group $U_{x,\mu}$ and $V_{x,\mu}$ according to the mid-point prescription, in view of the gauge transformations (\ref{U-transf}) for $U_{x,\mu}$ and (\ref{V-transf}) for $V_{x,\mu}$.
This inevitably leads to the $o(\epsilon^2)$ correction to the definition of $X_{x,\mu}$ in order to be consistent with 
$X_{x,\mu}=U_{x,\mu}V_{x,\mu}^{-1}=\exp \{ -i\epsilon g \mathscr{X}_\mu(x+\epsilon \hat\mu/2) 
+ \frac12 \epsilon^2 g^2[ \mathscr{A}_\mu(x+\epsilon \hat\mu/2), \mathscr{V}_\mu(x+\epsilon \hat\mu/2)] 
+o(\epsilon^3)  \}
$. 
}
\begin{align}
  X_{x,\mu} = \exp [-i\epsilon g \mathscr{X}_\mu(x) )]
 ,
 \quad
  V_{x,\mu} = \exp [-i\epsilon g \mathscr{V}_\mu(x+\epsilon \hat\mu/2)] 
  , \quad
\end{align}
just as the original link variable $U_{x,\mu}$ is associated with the gauge potential $\mathscr{A}_\mu(x)$: 
\begin{align}
U_{x,\mu} = \exp[ -i \epsilon g \mathscr{A}_\mu(x+\epsilon \hat\mu/2)]
 . 
\label{def-U}
\end{align}

Incidentally, we can consider another  decomposition of the form:
\begin{equation}
 U_{x,\mu} = \tilde{V}_{x,\mu} \tilde{X}_{x,\mu} 
  .
  \label{decomp-2}
\end{equation}
In this decomposition,  $\tilde{X}_{x,\mu}$ must obey the gauge transformation of the adjoint type at the site $x+\mu$:
\begin{equation}
  \tilde{X}_{x,\mu} (=  \tilde{V}_{x,\mu}^\dagger U_{x,\mu} ) \rightarrow \Omega_{x+\mu} \tilde{X}_{x,\mu} \Omega_{x+\mu}^\dagger  
  ,
  \label{X-transf2}
\end{equation}
provided that $U_{x,\mu}$ and $\tilde{V}_{x,\mu}$ transform in the same way as the above, (\ref{U-transf}) and (\ref{V-transf}) respectively. 
This case can be treated in the same way as the above case.
For instance, 
we can cast the relevant matrix element into the following form.
\begin{align}
\langle \xi_{x},\Lambda| U_{x,\mu}  | \xi_{x+\epsilon \hat\mu}, \Lambda \rangle
=& \langle \xi_{x},\Lambda| \tilde{V}_{x,\mu} \tilde{X}_{x,\mu} | \xi_{x+\epsilon \hat\mu}, \Lambda \rangle
\nonumber\\
=& \langle \Lambda| \xi_{x}^\dagger  \tilde{V}_{x,\mu} \xi_{x+\epsilon \hat\mu} \xi_{x+\epsilon \hat\mu}^\dagger \tilde{X}_{x,\mu} \xi_{x+\epsilon \hat\mu}  |  \Lambda \rangle
\nonumber\\
=&   (\xi_{x}^\dagger  \tilde{V}_{x,\mu} \xi_{x+\epsilon \hat\mu})_{dd} (\xi_{x+\epsilon \hat\mu}^\dagger  \tilde{X}_{x,\mu} \xi_{x+\epsilon \hat\mu})_{dd} 
\nonumber\\
=& \langle \xi_{x},\Lambda| \tilde{V}_{x,\mu} | \xi_{x+\epsilon \hat\mu}, \Lambda \rangle 
\times
\langle \xi_{x+\epsilon \hat\mu}, \Lambda| \tilde{X}_{x,\mu} | \xi_{x+\epsilon \hat\mu}, \Lambda \rangle  
 .
\end{align}
Therefore, the requirement (I) remains exactly the same, while the requirement (II) is replaced by a condition for 
$\langle \xi_{x+\epsilon \hat\mu}, \Lambda| \tilde{X}_{x,\mu} | \xi_{x+\epsilon \hat\mu}, \Lambda \rangle$. 
This fact implies that another decomposition (\ref{decomp-2}) gives the same $V_{x,\mu}$ as the above (\ref{decomp-1}), i.e.,  $\tilde{V}_{x,\mu}=V_{x,\mu}$.  
Although there exist some discrepancy between $X_{x,\mu}$ and $\tilde{X}_{x,\mu}$ at non-zero lattice spacing $\epsilon$ as suggested from the transformation rules (\ref{X-transf1}) and (\ref{X-transf2}), the difference between $X_{x,\mu}$ and $\tilde{X}_{x,\mu}$ vanishes in the continuum limit. 
Therefore, our results for the Abelian dominance does not depend on the way of the decomposition for the gauge variable.

In order to understand the requirement (I) more concretely,  we examine the SU(3) case. 
We adopt the Gell-Mann matrices $\lambda_A$ for the generators of su(3), i.e., $T_A=\lambda_A/2$ ($A=1, \cdots, 8$) where $\lambda_3$ and $\lambda_8$ are diagonal matrices.  Then it is shown \cite{Kondo08} that $\mathcal{H}=\lambda_8/(2\sqrt{3})$ is the simplest choice for $\mathcal{H}$ in fundamental representations of SU(3).  In this case, (\ref{req-Ia}) means that the Lie algebra of the group element $\xi_{x}^\dagger  V_{x,\mu} \xi_{x+\epsilon \hat\mu}$ is given as a linear combination of $\lambda_1$, $\lambda_2$, $\lambda_3$ and $\lambda_8$. 
This implies that $\hat{V}_{x,\mu}:=\xi_{x}^\dagger  V_{x,\mu} \xi_{x+\epsilon \hat\mu}$  is block diagonal and $\hat{V}_{x,\mu}  \in U(2)$ in agreement with $\tilde{H}=U(2)$.
From 
$\hat{U}_{x,\mu}:=\xi_{x}^\dagger U_{x,\mu}\xi_{x+\epsilon \hat\mu}
=(\xi_{x}^\dagger  X_{x,\mu} \xi_{x}) (\xi_{x}^\dagger  V_{x,\mu} \xi_{x+\epsilon \hat\mu})
=\hat{X}_{x,\mu}\hat{V}_{x,\mu}
$, we have  $\hat{X}_{x,\mu}=\hat{U}_{x,\mu}\hat{V}_{x,\mu}^{-1}$
and $\hat{X}_{x,\mu} \in SU(3)/U(2)$.
Therefore, the Lie algebra $\hat{\mathscr{X}}_{\mu}(x)$ of  $\hat{X}_{x,\mu}$  is written to be  a linear combination of  the matrices $\lambda_4$, $\lambda_5$, $\lambda_6$ and $\lambda_7$.

In the following, we examine how and when the requirements (I) and (II) are satisfied by clarifying their physical meanings. 
First of all, we observe that the condition (\ref{condition}) or the two requirements (I) and (II) are invariant under the lattice gauge transformations (\ref{V-transf}) and (\ref{X-transf1}), since $\xi_{x}^\dagger  V_{x,\mu} \xi_{x+\epsilon \hat\mu}$ and $\xi_{x}^\dagger  X_{x,\mu} \xi_{x}$ are gauge invariant combinations.  
Therefore, they could hold exactly on a lattice for arbitrary $\epsilon$. 
However, we show later that the requirement (II) (\ref{req-II}) is exactly satisfied only in the continuum limit $\epsilon \rightarrow 0$, while the requirement (I) is fulfilled for arbitrary lattice spacing $\epsilon$. We shall see that the constant in (\ref{req-II}) goes to one in the naive continuum limit $\epsilon \rightarrow 0$.


It should be remarked that the Wilson criterion for quark confinement is not applicable at finite temperature.  A criterion for confinement at finite temperature is given in terms of the Polyakov  loop operator $L_{\vec x}[U]$ (or the thermal Wilson line) which goes along the temporal direction \cite{Polyakov78}: 
\begin{equation}
 L_{\vec x}[U] = {\rm tr}(\mathscr{P} \prod_{x_D}  U_{x,\mu=D} )/{\rm tr}(\bm{1}) , \quad x=(\vec{x},x_D)
  .
\end{equation}
If the conditions (\ref{condition}) is satisfied, the ``Abelian dominance'' in the same sense as the Wilson loop holds: 
\begin{equation}
 L_{\vec x}[U] =  {\rm const.} L_{\vec x}[V] 
  .
\end{equation}

\section{Formulae for the coherent states}

In order to see the physical meaning of the requirements (I) and(II),  we prepare some formulae for the coherent state which are needed to rewrite the requirements.
 
For the Wilson loop operator in the fundamental representations, the matrix element of any Lie algebra valued operator $ \mathscr{O}$ in the coherent state is cast into 
\begin{subequations}
\begin{align}
 \langle \xi,\Lambda| \mathscr{O}    | \tilde\xi, \Lambda \rangle 
=&  \langle \Lambda| \xi^\dagger  \mathscr{O}  \tilde\xi |  \Lambda \rangle 
\nonumber\\
=& (\xi^\dagger  \mathscr{O}  \tilde\xi )_{dd} \ (\text{no sum over $d$})
\label{dd-element}
\\
=& {\rm tr}[ \xi^\dagger \mathscr{O}  \tilde\xi  \bm{e}_{dd}    ]
\\
=& {\rm tr}[ \mathscr{O}  \tilde\xi  \bm{e}_{dd} \xi^\dagger   ]
 .
\end{align}
\end{subequations}
where we have introduced a matrix $\bm{e}_{dd}$ having only one  non-vanishing  diagonal element: $(\bm{e}_{dd})_{bc}=\delta_{db}\delta_{dc}$ (no sum over $d$).

We can introduce $\mathcal{H}$ given by \cite{Kondo08}
\begin{equation}
 \mathcal{H} := 
\bm{\Lambda} \cdot \bm{H}  = \sum_{j=1}^{r} \Lambda_j H_j 
 ,
\end{equation}
where  $H_j$ ($j=1,, \cdots, r=N-1$) are the generators from the Cartan subalgebra ($r=N-1$ is the rank of the gauge group $G=SU(N)$) and 
$r$-dimensional vector $\Lambda_j$ ($j=1,, \cdots, r$) is the highest weight of the representation in which the Wilson loop is considered. 
For $N$ fundamental representations for SU(N) ($d=1, \cdots, N$),  we have 
\begin{equation}
  \bm{e}_{dd} := \frac{1}{{\rm tr}(\bm{1})} \bm{1} + 2 \mathcal{H}  
  , \quad
  \mathcal{H}  = \frac12 \left( \bm{e}_{dd} - \frac{1}{{\rm tr}(\bm{1})} \bm{1}  \right) 
   .
\end{equation}
Hence the  matrix element of any Lie algebra valued operator $ \mathscr{O}$  in the coherent state \cite{Kondo08} reads
\begin{subequations}
\begin{align}
  \langle \xi,\Lambda| \mathscr{O}    | \tilde\xi, \Lambda \rangle 
=& {\rm tr}[ \mathscr{O}  \tilde\xi  \bm{e} \xi^\dagger   ]
\nonumber\\
=& {\rm tr} \left[ \xi^\dagger \mathscr{O} \tilde\xi  \left( \frac{1}{{\rm tr}(\bm{1})} \bm{1}   + 2 \mathcal{H} \right)   \right]
\label{xOx}
\\
=& {\rm tr} \left[ \mathscr{O}   \left( \frac{1}{{\rm tr}(\bm{1})} \tilde\xi \xi^\dagger  + 2 \tilde\xi \mathcal{H} \xi^\dagger   \right)   \right]
 .
 \label{matrix-element}
\end{align}
\end{subequations}

We  introduce  a new field $\bm{m}(x)$ having its value in the Lie algebra $\mathscr{G}=su(N)$ by 
\begin{equation}
  \bm{m}(x)  := \xi(x) \mathcal{H} \xi(x)^\dagger 
   =  \sum_{j=1}^{r}  \Lambda_j \bm{n}_j(x) 
    ,
    \label{m-n}
\end{equation}
which is a linear combination of the color field defined by
\begin{equation}
 \bm{n}_j(x)   :=    \xi(x)H_j  \xi(x)^\dagger \quad
 (j=1, \cdots, N-1)
    .
\end{equation}
Moreover, we can introduce the normalized color field $\bm{n}(x)$ by
\begin{equation}
  \bm{n}(x)  
  := \sqrt{\frac{2N}{N-1}} \bm{m}(x) 
  = \sqrt{\frac{2N}{N-1}}  \xi(x) \mathcal{H} \xi(x)^\dagger 
\label{unit-n}
    .
\end{equation}
In particular, the diagonal matrix element is cast into \cite{Kondo08} 
\begin{align} 
  \langle \xi(x),\Lambda|\mathscr{O}(x)  | \xi(x), \Lambda \rangle 
=   {\rm tr} \left\{ \mathscr{O}(x)  \left[ \frac{1}{{\rm tr}(\bm{1})}\bm{1} + 2 \bm{m}(x) \right]   \right\} 
 ,
 \label{matrix-element-2}
\end{align}
and
\begin{align} 
  \langle \Lambda|\mathscr{O}(x)  |   \Lambda \rangle 
=   {\rm tr} \left\{  \left[ \frac{1}{{\rm tr}(\bm{1})}\bm{1} + 2 \mathcal{H} \right]  \mathscr{O}(x)  \right\} 
 .
\end{align}

\section{Field decomposition and Abelian dominance}

For SU(2), $\bm{m}(x)$ agrees with the unit color field $\bm{n}(x)$ introduced in the previous paper \cite{KMS06}:
\begin{equation}
\bm{m}(x) = \frac12 \bm{n}(x)
 .
 \label{SU2}
\end{equation}
The color field $\bm{n}(x)$ denotes a spacetime-dependent embedding of the Abelian direction into the non-Abelian color space and hence the Abelian direction can vary from point to point of  spacetime. 
Therefore, the color field $\bm{n}(x)$ plays the role of recovering color symmetry which will be lost by a global (i.e., space-time independent or uniform) choice of the Abelian direction taken in the conventional approach, e.g., the maximal Abelian gauge. 
The field strength $\mathscr{F}_{\mu\nu}^{[\mathscr{V}]}(x)$ of the non-Abelian gauge field $\mathscr{V}_\mu(x)$ at $x$ is in general 
non-Abelian and is not necessarily proportional to the Abelian direction at $x$ specified by the color field $\bm{n}(x)$ at $x$.

In view of this, we consider the meaning of the first requirement (I):  
\begin{align}
  \xi_{x}^\dagger  V_{x,\mu} \xi_{x+\epsilon \hat\mu}  \in \tilde{H} 
  ,
\end{align}
which is equivalent to 
\begin{subequations}
\begin{align} 
 [ \mathcal{H}, \xi_{x}^\dagger  V_{x,\mu} \xi_{x+\epsilon \hat\mu} ] = 0 
  \label{req-Ia}
  ,
\end{align}
and
\begin{align} 
 \bm{m}_{x} V_{x,\mu} = V_{x,\mu} \bm{m}_{x+\mu} 
  ,
  \label{req-Ib}
\end{align}
\end{subequations}
where (\ref{req-Ib}) is obtained by multiplying both sides of (\ref{req-Ia})  by $\xi_{x}$ from the left and $\xi_{x+\mu}^\dagger$ from the right.  
The equation (\ref{req-Ib}) agrees with the first defining equation for the decomposition (\ref{decomp-1}) on a lattice derived in \cite{IKKMSS06,SKKMSI07} for SU(2) and \cite{KSSMKI08,SKKMSI07b} for SU(N).

Now we show that the requirement (I), (\ref{req-Ia}) or (\ref{req-Ib}), is a condition for the field strength $\mathscr{F}_{\mu\nu}^{[\mathscr{V}]}(x)$ of the restricted  field $\mathscr{V}_\mu(x)$ to have only the Abelian part  proportional to the color field $\bm{n}(x)$ at the same spacetime point $x$.

On a lattice, the arguments are as follows. 
We recall a general relation in lattice gauge theory: the path-ordered product of the gauge variables along the four links around a plaquette (elementary square) is related to the field strength of the gauge field: 
\begin{align} 
 V_{x,\mu}V_{x+\mu,\nu}
V_{x+\nu,\mu}^{\dagger}V_{x,\nu}^{\dagger} 
= \exp \left[ -i \epsilon^2 g \mathscr{F}_{\tilde{x},\mu\nu}^{[\mathscr{V}]}  + o(\epsilon^4) \right] 
 ,
 \label{plaquette}
\end{align}
where $\tilde{x}$ denotes 
$\tilde{x}:=x+\epsilon \hat\mu/2+\epsilon \hat\nu/2$.
It is easy to see that the commutativity of the plaquette variable $V_{x,\mu}V_{x+\hat{\mu},\nu}
V_{x+\nu,\mu}^{\dagger}V_{x,\nu}^{\dagger}$ with the color field $\bm{n}_{x}$ at the corner $x$ of the plaquette  on a lattice 
\begin{align} 
 [ V_{x,\mu}V_{x+\mu,\nu}
V_{x+\nu,\mu}^{\dagger}V_{x,\nu}^{\dagger} , \bm{m}_{x} ]
= 0 
 ,
 \label{F-commu}
\end{align}
follows from a more elementary relation (\ref{req-Ib}). 
The relation (\ref{req-Ib}) means that the color field $\bm{m}_{x}$ at a site $x$ can be parallel transported to the next site $x+\mu$ across the link  $<x,x+\mu>$ in the background $V_{x,\mu}$ and is identified with $\bm{m}_{x+\mu}$ at the next cite.

For SU(2), the commutativity (\ref{F-commu}) means that the plaquette variable (\ref{plaquette}) is written in the form 
\begin{align} 
 V_{x,\mu}V_{x+\mu,\nu}
V_{x+\nu,\mu}^{\dagger}V_{x,\nu}^{\dagger} 
= \exp \left[ -i \epsilon^2 g F_{\tilde{x},\mu\nu}  \bm{n}_{\tilde{x}} + o(\epsilon^3) \right] 
 ,
\end{align}
where $F_{\tilde{x},\mu\nu}$ is the SU(2) gauge invariant field strength in the Abelian direction specified by $\bm{n}_{\tilde{x}}$ at  $\tilde{x}$. 
Thus we conclude that 
$
\mathscr{F}_{\tilde{x},\mu\nu}^{[\mathscr{V}]}=F_{\tilde{x},\mu\nu}  \bm{n}_{\tilde{x}}
$
for SU(2).
The case of SU(N) ($N \ge 3$) can be treated in the similar way to the SU(2) case. See \cite{Kondo08} for details.

In the continuum theory,  the commutator between   $\mathscr{F}_{\mu\nu}^{[\mathscr{V}]}(x)$  and a color field $\bm{m}(x)$ is calculated using 
$
 \mathscr{F}_{\mu\nu}^{[\mathscr{V}]} 
= ig^{-1} [ D_\mu^{[\mathscr{V}]}, D_\nu^{[\mathscr{V}]} ]
$
for 
$
 D_\mu^{[\mathscr{V}]} := \partial_\mu -ig [ \mathscr{V}_\mu , \cdot ] 
$ 
as follows.
\begin{align}
 [ \mathscr{F}_{\mu\nu}^{[\mathscr{V}]}, \bm{m} ]
 =& ig^{-1}  [ [ D_\mu^{[\mathscr{V}]}, D_\nu^{[\mathscr{V}]} ], \bm{m}  ] 
 \nonumber\\
 =&  - ig^{-1} [ [ D_\nu^{[\mathscr{V}]} , \bm{m} ], D_\mu^{[\mathscr{V}]}]  - ig^{-1} [ [ \bm{m}, D_\mu^{[\mathscr{V}]} ], D_\nu^{[\mathscr{V}]}]  
 \nonumber\\
 =&   ig^{-1} [ D_\mu^{[\mathscr{V}]},  (D_\nu^{[\mathscr{V}]}  \bm{m}) ]  + ig^{-1} [  D_\nu^{[\mathscr{V}]}, (D_\mu^{[\mathscr{V}]} \bm{m})  ]  
 \nonumber\\
 =&   ig^{-1} [ D_\mu^{[\mathscr{V}]},  D_\nu^{[\mathscr{V}]} ] \bm{m}  
 ,
\end{align}
where we have used the Jacobi identity in the second equality 
and the relation:
$
 [ D_\mu^{[\mathscr{V}]} , \bm{m} ] f
  = (D_\mu^{[\mathscr{V}]}  \bm{m}) f 
$
for an arbitrary function $f$ in the third and the last equalities.
Thus, if  $\mathscr{V}_\mu(x)$ and $\bm{m}(x)$ satisfies%
\footnote{
It is shown that (\ref{req-Ib}) is a lattice version of (\ref{DVm}).   
In fact, expanding (\ref{req-Ib}) around the mid-point $x+\epsilon \hat\mu/2$ yields (\ref{DVm}) to the accuracy of the next order $\epsilon^2$, which is to be compared with the naive expansion around  end-points. 
See \cite{SKKMSI07b} and \cite{KSSMKI08} for more details. 
}
\begin{equation}
 D_\mu^{[\mathscr{V}]}  \bm{m}(x) = 0
    ,
    \label{DVm}
\end{equation}
then $\mathscr{F}_{\mu\nu}^{[\mathscr{V}]}(x)$ and $\bm{m}(x)$ commute, i.e., 
\begin{equation}
[ \mathscr{F}_{\mu\nu}^{[\mathscr{V}]}(x) , \bm{m}(x) ]
= 0 
 .
 \label{F-m}
\end{equation}
For SU(2), this means that the field strength $\mathscr{F}_{\mu\nu}^{[\mathscr{V}]}(x)$ of the field $\mathscr{V}_\mu(x)$ has only the Abelian part proportional to $\bm{n}(x)$: 
\begin{equation}
 \mathscr{F}_{\mu\nu}^{[\mathscr{V}]}(x) =   F_{\mu\nu}(x) \bm{n}(x) 
    ,
    \label{F=Fn}
\end{equation}
since $\mathscr{F}_{\mu\nu}^{[\mathscr{V}]}(x)$ is traceless and can not have a part proportional to the unit matrix.  
This is the stronger sense of ``Abelian dominance''.
Moreover, the condition (\ref{DVm}) agrees with the first  defining equation for the CFNS decomposition 
$\mathscr{A}_\mu(x)=\mathscr{V}_\mu(x)+\mathscr{X}_\mu(x)$ in the case of SU(2).

The above argument can be extended to SU(N) case, which is more involved from the technical points of view.  
See \cite{Kondo08} and also \cite{KSM08} for the full details. 
The relations (\ref{req-Ib}) and (\ref{F-commu}) are actually a lattice version of the continuum relations (\ref{DVm}) and (\ref{F-m}).

 The equation obtained from (\ref{xOx}): 
\begin{align}
  \langle \xi_{x},\Lambda| V_{x,\mu} | \xi_{x+\epsilon \hat\mu}, \Lambda \rangle 
=& {\rm tr} \left[ \xi_{x}^\dagger V_{x,\mu}  \xi_{x+\epsilon \hat\mu}       \right]/{\rm tr}(\bm{1})  
+ 2{\rm tr} \left[ \xi_{x}^\dagger V_{x,\mu}  \xi_{x+\epsilon \hat\mu}    \mathcal{H}  \right]
 .
\end{align}
shows  that the requirement (I) (\ref{req-Ia}) is automatically incorporated in this approach. 
Therefore, once the Wilson loop operator is rewritten into the surface integral form through the  non-Abelian Stokes theorem using the coherent state, it is represented by the (SU(N) gauge-invariant) Abelian field strength  $F$, i.e., the magnitude of the field strength $\mathscr{F}$ of the restricted field $\mathscr{V}$ \cite{Kondo08}:
\begin{equation}
W_C[\mathscr{V}]
=W_S[F]
:=  \prod_{x \in S} \int d\mu(\xi_{x}) \exp \left( ig \  \int_{S:\partial S=C} F  \right) ,
\quad 
F:= 2{\rm tr}(\bm{m}(x)\mathscr{F}[\mathscr{V}](x))
    .
    \label{NAST2}
\end{equation} 
Thus we have obtained a gauge-invariant (and gauge-independent) definition of the Abelian dominance in the continuum form: 
\begin{equation}
W_C[\mathscr{A}]= {\rm const.} W_C[\mathscr{V}]= {\rm const.} W_S[F]
    ,
\end{equation} 
where $\mathscr{V}$ is defined in a gauge-covariant and gauge-independent way and $F$ becomes gauge invariant and hence gauge independent.

Now we turn our attention to the physical meaning of the second requirement (II).  
We decompose the matrix element $ \langle \xi_{x}, \Lambda| X_{x,\mu} | \xi_{x}, \Lambda \rangle$ into the real and imaginary parts: 
\begin{align}
 \langle \xi_{x}, \Lambda| X_{x,\mu} | \xi_{x}, \Lambda \rangle
=& Re\langle \xi_{x}, \Lambda| X_{x,\mu} | \xi_{x}, \Lambda \rangle + i Im \langle \xi_{x}, \Lambda| X_{x,\mu} | \xi_{x}, \Lambda \rangle
\nonumber\\
=&  \langle \xi_{x}, \Lambda| \frac{X_{x,\mu}+X_{x,\mu}^\dagger}{2} | \xi_{x}, \Lambda \rangle 
+   \langle \xi_{x}, \Lambda| \frac{X_{x,\mu}-X_{x,\mu}^\dagger}{2} | \xi_{x}, \Lambda \rangle 
 .
 \label{X-1}
\end{align}
This is understood by applying the formula (\ref{matrix-element-2}) to the two terms on the right-hand side:
\begin{align} 
 & \langle \xi_{x}, \Lambda| \frac{X_{x,\mu} \pm X_{x,\mu}^\dagger}{2} | \xi_{x}, \Lambda \rangle 
\nonumber\\
=&   {\rm tr} \left\{ \frac{X_{x,\mu} \pm X_{x,\mu}^\dagger}{2}   \right\} /{\rm tr}(\bm{1})
+ {\rm tr} \left\{  ( X_{x,\mu} \pm X_{x,\mu}^\dagger )    \bm{m}_{x}    \right\} 
 .
 \label{xi-xi}
\end{align}
The matrix element $ \langle \xi_{x}, \Lambda| X_{x,\mu} | \xi_{x}, \Lambda \rangle$ can not be a real number, since the imaginary part can not vanish except for the limit $\epsilon \rightarrow 0$. 
This is easily seen by estimating the order in the lattice spacing $\epsilon$ of the four terms by expanding the $X_{x,\mu}$ and $X_{x,\mu}^\dagger$in power series in $\epsilon$:
$
X_{x,\mu}$
or 
$X_{x,\mu}^\dagger$
$=1 \mp i \epsilon g\mathscr{X}_\mu(x) - \frac12 \epsilon^2 g^2 \mathscr{X}_\mu(x)^2 + o(\epsilon^3)
$ as%
\footnote{
The reason that the $o(\epsilon)$ term vanishes in ${\rm tr} \left\{( X_{x,\mu} - X_{x,\mu}^\dagger)/2 \right\}$ comes from the fact ${\rm tr}(\mathscr{X}_\mu(x))=0$ due to a property of the su(N) Lie algebra for an SU(N) group element $X_{x,\mu}$. This is not the case for u(N) Lie algebra. 
}
\begin{align} 
 &{\rm tr} \left\{ \frac{X_{x,\mu} + X_{x,\mu}^\dagger}{2}   \right\}/{\rm tr}(\bm{1}) 
= 1 + O(\epsilon^2)
  ,
\nonumber\\
&   {\rm tr} \left\{  ( X_{x,\mu} + X_{x,\mu}^\dagger )    \bm{m}_{x}    \right\} 
= O(\epsilon^2) 
  ,
\nonumber\\
 &{\rm tr} \left\{ \frac{X_{x,\mu} - X_{x,\mu}^\dagger}{2}   \right\}/{\rm tr}(\bm{1}) 
= O(\epsilon^3)
  ,
\nonumber\\
&   {\rm tr} \left\{  ( X_{x,\mu} - X_{x,\mu}^\dagger )    \bm{m}_{x}    \right\} 
= -ig\epsilon 2{\rm tr}(\mathscr{X}_{x,\mu} \bm{m}_{x}) +  O(\epsilon^3) 
= O(\epsilon^1) 
 .
 \label{xi-xi-2}
\end{align}
The order $\epsilon^2$ and higher-order terms are usually neglected to obtain the path-integral representation, since the path-integral is one-dimensional line integral to keep only the order $\epsilon^1$ terms which survive in the continuum limit. 
Therefore, the second and third terms in (\ref{xi-xi-2}) are neglected. 
In view of the continuum limit, we impose the condition  for the Lie algebra:
\begin{equation}
{\rm tr}(\mathscr{X}_{x,\mu} \bm{m}_{x}) = 0 
  ,
\label{Xm}
\end{equation}
which implies the condition for matrices $X_{x,\mu}$ as group elements up to $O(\epsilon^3)$:
\begin{equation}
 {\rm tr} \left\{  ( X_{x,\mu} - X_{x,\mu}^\dagger )  \bm{m}_{x}    \right\} 
 = 0 
 .
 \label{req-IIc}
\end{equation}
By imposing (\ref{Xm}), therefore, the matrix element $\langle \xi_{x}, \Lambda| X_{x,\mu} | \xi_{x}, \Lambda \rangle$ has a real value up to $O(\epsilon^2)$ and the requirement (II) is fulfilled: 
\begin{equation}
 \langle \xi_{x}, \Lambda| X_{x,\mu} | \xi_{x}, \Lambda \rangle
=  1 + O(\epsilon^2) 
 .
\end{equation}

The condition (\ref{Xm}) agrees with the second defining equation of the decomposition  for the traceless su(N) Lie algebra valued field $\mathscr{X}_\mu$:
\begin{align} 
 0   
=    {\rm tr} \left\{ \mathscr{X}_\mu(x)   \bm{m}(x)  
\right\}
 ,
\label{condition-2}
\end{align}
which is equivalent to 
${\rm tr} \left\{ \hat{\mathscr{X}}_{\mu}(x) \mathcal{H}   \right\} 
=0$ for the Lie algebra $\hat{\mathscr{X}}_{\mu}(x)$ of $\hat{X}_{x,\mu}$. 
The equation (\ref{req-IIc}) is a modified form of the second defining equation 
$
{\rm tr} \left\{  X_{x,\mu}   \bm{m}_{x}    \right\} 
=    O(\epsilon^2)
$
for the decomposition (\ref{decomp-1}) on a lattice derived in \cite{SKKMSI07} for SU(2) and \cite{KSSMKI08} for SU(N).

Thus, in the continuum limit,  we have obtained the dominance of the decomposed variable $V_{x,\mu}$ in the SU(N) Wilson loop operator by requiring no contributions (\ref{Xm}) from the variable $X_{x,\mu}$ to the Wilson loop operator, which is regarded as the ``Abelian dominance'' by abuse of language, see \cite{Kondo08} for more details.

For the SU(2) case, it was already pointed out by Cho \cite{Cho00} that the variable $\mathscr{V}_\mu(x)$ as one of the decomposed variables of the original gauge potential $\mathscr{A}_\mu(x)$ according to the CFNS decomposition $\mathscr{A}_\mu(x)=\mathscr{V}_\mu(x)+\mathscr{X}_\mu(x)$
 is dominant in the Wilson loop within the continuum formulation.
In fact,   (\ref{condition-2}) for SU(2), i.e., 
\begin{align} 
 0 =  {\rm tr} \left\{ \mathscr{X}_\mu(x)  \bm{n}(x)   
\right\}
 ,
\label{def1}
\end{align}
agrees with one of the two defining equations which are able to uniquely determine the CFNS decomposition.

If the color field is fixed to be uniform
\begin{equation}
\bm{n}(x)=\sigma_3/2
 , 
\end{equation}
the requirement (\ref{def1}) reduces to
\begin{equation}
 0 =  {\rm tr} \left\{ \mathscr{X}_\mu(x)    \sigma_3 
\right\}
 ,
\end{equation}
which implies that $\mathscr{X}_\mu(x)$ is the off-diagonal matrix.  Then the restricted field $\mathscr{V}_\mu(x)$ becomes the diagonal component 
$
\mathscr{V}_\mu(x) = \mathscr{A}_\mu^3 \frac{\sigma_3}{2}
$. 
Consequently, the requirement (\ref{DVm}): 
\begin{equation}
 0 = D_\mu^{[\mathscr{V}]}  \bm{n}(x) := \partial_\mu \bm{n}(x)-ig [\mathscr{V}_\mu(x) , \bm{n}(x)] 
    ,
    \label{DVn}
\end{equation}
is automatically satisfied. 
Thus,  $\mathscr{A}_\mu
= \mathscr{V}_\mu(x)  + \mathscr{X}_\mu(x)$ reduces just to  the decomposition into the diagonal and off-diagonal components:
\begin{equation}
\mathscr{A}_\mu(x)
= \mathscr{A}_\mu^A(x) \frac{\sigma_A(x)}{2} \ (A=1,2,3) ,
\ 
\mathscr{V}_\mu(x) = \mathscr{A}_\mu^3(x) \frac{\sigma_3}{2} ,
\
\mathscr{X}_\mu(x) = \mathscr{A}_\mu^a(x) \frac{\sigma_a}{2}  \ (a=1,2) 
 .
\end{equation}

\section{Deviation from the exact Abelian dominance}

Finally, we estimate the deviation on a lattice from the exact (or 100$\%$) Abelian dominance which is realized in the continuum limit. 
Under the requirement (I), the difference between the full Wilson loop $W_C[U]$ and the restricted Wilson loop $W_C[V]$ reads
\begin{align}
& W_C[V]-W_C[U] 
\nonumber\\ 
=& \prod_{x \in C} \int d\mu(\xi_x) 
\prod_{\ell=<x,x+\epsilon \hat\mu> \in C} \langle \xi_{x},\Lambda| V_{x,\mu} | \xi_{x+\epsilon \hat\mu}, \Lambda \rangle 
 R_C[X,\xi]
  ,
\end{align} 
where we have introduced
\begin{align}
R_C[X,\xi]
:=& 1 - \prod_{<x,x+\mu> \in C} \langle \xi_{x}, \Lambda| X_{x,\mu} | \xi_{x}, \Lambda \rangle   
\nonumber\\
=& 1 - \exp \{ \sum_{<x,x+\mu> \in C} \ln \langle \xi_{x}, \Lambda| X_{x,\mu} | \xi_{x}, \Lambda \rangle \}  
 .
\end{align}
Therefore, the deviation 
$|W_C[V]-W_C[U]|$ is bounded from the above:
\begin{align}
& |W_C[V]-W_C[U]| 
\nonumber\\ 
=& | \prod_{x \in C} \int d\mu(\xi_x) 
\prod_{\ell=<x,x+\epsilon \hat\mu> \in C} \langle \xi_{x},\Lambda| V_{x,\mu} | \xi_{x+\epsilon \hat\mu}, \Lambda \rangle 
 R_C[X,\xi]|
\nonumber\\ 
\le &  \prod_{x \in C} \int d\mu(\xi_x) |R_C[X,\xi]|
  .
\end{align}

For concreteness, we estimate the matrix element up to $O(\epsilon^2)$ by imposing (\ref{Xm}). Note that the $O(\epsilon^2)$ contributions come from the real part of $ \langle \xi_{x}, \Lambda| X_{x,\mu} | \xi_{x}, \Lambda \rangle$, since the pure imaginary part in (\ref{xi-xi-2}) becomes $O(\epsilon^3)$ after imposing (\ref{Xm}) as shown in the above.  Hence the real part in   (\ref{xi-xi-2}) reads
\begin{align} 
& Re \langle \xi_{x}, \Lambda| X_{x,\mu} | \xi_{x}, \Lambda \rangle
\nonumber\\
=& \langle \xi_{x}, \Lambda| \frac{X_{x,\mu}+X_{x,\mu}^\dagger}{2} | \xi_{x}, \Lambda \rangle
\nonumber\\
=&   {\rm tr} \left[ \frac{X_{x,\mu} + X_{x,\mu}^\dagger}{2}   \right]/{\rm tr}(\bm{1}) 
+   {\rm tr} \left[  ( X_{x,\mu} + X_{x,\mu}^\dagger )    \bm{m}_{x}    \right]
\nonumber\\
=&   1 - \frac12 g^2 \epsilon^2 {\rm tr} [\mathbb{X}_\mu(x)^2 ]/{\rm tr}(\bm{1}) 
- \frac12 g^2 \epsilon^2 {\rm tr} [\mathbb{X}_\mu(x)^2 \bm{m}(x)] 
+  O(\epsilon^3)
  .
 \label{xi-xi-3}
\end{align}
Here note that 
$
{\rm tr} [\mathbb{X}_\mu(x)^2 ]
= \frac12 \mathbf{X}_\mu \cdot \mathbf{X}_\mu
:=  \frac12 X_\mu^A  X_\mu^A 
$ 
($A=1, \cdots, N^2-1$)
  is positive definite for Euclidean space,
while
${\rm tr} [\mathbb{X}_\mu(x)^2 \bm{m}(x)]
 = \frac14 (\mathbf{X}_\mu * \mathbf{X}_\mu ) \cdot \mathbf{m}
 :=   \frac14   d^{ABC} X_\mu^A X_\mu^B m^C
$  
 is zero for SU(2) and non-trivial only for SU($N$) ($N \ge 3$) without positive definiteness.%
\footnote{
We have used 
$  {\rm tr}(T^A T^B T^C ) =   {1 \over 4} (d^{ABC}+if^{ABC})  
$,
which follows from the basic relation:
$  T^A T^B = {1 \over 2N} \delta^{AB} \bm{1} + {1 \over 2} (d^{ABD}+if^{ABD}) T^D
$.

}

For simplicity, we focus on the SU(2) case:
\begin{align} 
R_C[X,\xi] 
=&  1-  \exp \left\{ -\frac12 g^2 \epsilon^2 \sum_{<x,x+\mu> \in C} ( {\rm tr} [\mathbb{X}_\mu(x)^2 ]/{\rm tr}(\bm{1}) 
+  O(\epsilon^2) ) \right\}   
 .
 \label{prod-xi-xi}
\end{align}
Then we obtain the bound up to $O(\epsilon^2)$: 
\begin{align}
 |W_C[V]-W_C[U]| 
\le &   \frac12 g^2 \epsilon^2 \sum_{<x,x+\mu> \in C} ( {\rm tr} [\mathbb{X}_\mu(x)^2 ]/{\rm tr}(\bm{1}) 
+  O(\epsilon^2) )
 .
 \label{error}
\end{align}

For a fixed $\epsilon>0$, the deviation has the smaller value for the smaller value of
$
 \sum_{<x,x+\mu> \in C}  {\rm tr} [\mathbb{X}_\mu(x)^2 ]
$.
Thus we have given a formula (\ref{error}) estimating a systematic error of the lattice Wilson loop operator from the continuum Wilson loop operator due to the non-zero lattice spacing $\epsilon$.

\section{The conventional MAG and Abelian dominance}

By imposing the requirement (I), the Wilson loop operator reads 
\begin{align}
 W_C[U]  
=&  \int \prod_{x}d\mu(\xi_x) 
\rho_C[X,\xi] w_C[V,\xi] 
 ,
 \label{afterI}
\end{align} 
where we have extended the region of integration for $\xi_{x}$ to the whole spacetime $x \in \mathbb{R}^D$ and defined
\begin{align}
  \rho_C[X,\xi]:=& \prod_{\ell=<x,x+\epsilon \hat\mu> \in C} \langle \xi_{x}, \Lambda| X_{x,\mu} | \xi_{x}, \Lambda \rangle  
 ,
\nonumber\\
  w_C[V,\xi]:=& \prod_{\ell=<x,x+\epsilon \hat\mu> \in C} 
\langle \xi_{x},\Lambda| V_{x,\mu} | \xi_{x+\epsilon \hat\mu}, \Lambda \rangle 
 .
 \label{degI}
\end{align} 
We can see from (\ref{X-1})  and  (\ref{xi-xi}) that  $\rho_C[X,\xi]$ is rewritten in terms of $X_{x,\mu}$ and $\bm{m}_{x}$, i.e., $\rho_C[X,\xi]=\tilde\rho_C[X,\bm{m}]$.  In the similar way, it is shown that $w_C[V,\xi]$ is also rewritten in terms of $V_{x,\mu}$ and $\bm{m}_{x}$ \cite{Kondo08},  i.e., $w_C[V,\xi]=\tilde{w}_C[V,\bm{m}]$.
In other words, $\{ \xi_{x} \}$ appears only in the combination $\{ \xi_{x} \mathcal{H} \xi_{x}^\dagger \}$ in the integrand of (\ref{afterI}). 
By inserting the unity:
\begin{equation}
 1 = \prod_{x} \int d\bm{m}_{x} \delta( \bm{m}_{x}- \xi_{x} \mathcal{H} \xi_{x}^\dagger) 
  ,
\end{equation}
into the integrand of (\ref{afterI}), therefore, the Wilson loop operator is cast into 
\begin{align}
 W_C[U]  
=&  \int \prod_{x} d\bm{m}_{x}  \int \prod_{x}d\mu(\xi_x) \prod_{x} \delta( \bm{m}_{x}- \xi_{x} \mathcal{H} \xi_{x}^\dagger) 
\rho_C[X,\xi] w_C[V,\xi] 
\nonumber\\
=& \int \prod_{x} d\bm{m}_{x}  
 \tilde\rho_C[X,\bm{m}] \tilde{w}_C[V,\bm{m}]
 ,
 \label{afterI2}
\end{align} 
where the last factor is given from (\ref{xi-xi-3})
\begin{align} 
\tilde\rho_C[X,\bm{m}]
=&  \exp \left\{ -\frac{g^2 \epsilon}{2} \epsilon \sum_{<x,x+\mu> \in C} \left(  {\rm tr} [\mathbb{X}_\mu(x)^2 ]/{\rm tr}(\bm{1}) 
+ {\rm tr} [\mathbb{X}_\mu(x)^2 \bm{m}(x)]   
+  O(\epsilon^2) \right) \right\}   
 .
 \label{rho_C}
\end{align}
Remember that $\mathbb{X}_\mu(x)$ and $\mathbb{V}_\mu(x)$ are written in terms of $\mathscr{A}_\mu(x)$ and $\bm{m}(x)$  by solving the defining equations for the decomposition, i.e., $\mathbb{X}=\mathbb{X}[\mathscr{A},\bm{m}]$ and $\mathbb{V}=\mathbb{V}[\mathscr{A},\bm{m}]$.
For instance, we have for SU(N)
\begin{equation}
 {\rm tr} [\mathbb{X}_\mu(x)^2 ]
 = \frac{2(N-1)}{Ng^2} {\rm tr} [(D_\mu[\mathscr{A}] \bm{m})^2]
  .
\end{equation}

In order to estimate how the integrand behaves with respect to the integration variables $\bm{m}(x)$, 
 we consider the Wilson loop operator $W_C[U]$ for a given configuration of $U_{x,\mu}$ (or $\mathscr{A}_\mu(x)$).  
For this purpose, we rewrite $\tilde{w}_C[V,\bm{m}]=\tilde{w}_C[V[U,\bm{m}],\bm{m}]=\hat{w}_C[U,\bm{m}]$ and 
$\tilde\rho_C[X,\bm{m}]=\tilde\rho_C[X[U,\bm{m}],\bm{m}]=\hat\rho_C[U,\bm{m}]$ to obtain 
\begin{align}
 W_C[U]  
= \int \prod_{x} d\bm{m}(x)   
\hat\rho_C[U,\bm{m}]  \hat{w}_C[U,\bm{m}]
 .
 \label{afterI3}
\end{align} 
It is not difficult to show that the functional $\hat\rho_C[U,\bm{m}]$ of $\{ \bm{m}(x)  \}$ and $\{ U_{x,\mu} \}$ is positive and is bounded from the above, 
\begin{equation}
0<\hat\rho_C[U,\bm{m}]= \tilde\rho_C[X,\bm{m}]<1 
  .
\end{equation} 
In particular, for SU(2), the integrand of the exponential in $\hat\rho_C[U,\bm{m}]$ is quadratic in $\bm{m}$:
\begin{align} 
\hat\rho_C[U,\bm{m}]
= \tilde\rho_C[X,\bm{m}] 
=   \exp \left\{ -\frac{\epsilon^2}{4} \sum_{<x,x+\mu> \in C}  ( {\rm tr} [(D_\mu[\mathscr{A}] \bm{m})^2]
+  O(\epsilon^2) )  \right\}   
 .
 \label{rho_C2}
\end{align}

An approximate Abelian dominance follows if the following situation occurs: if we regard $\prod_{x} d\bm{m}(x)   
\hat\rho_C[U,\bm{m}]$ as  the integration measure and  $\hat{w}_C[U,\bm{m}]$ as the integrand of $W_C[U]$ in (\ref{afterI3}), 
 the integral over all possible configurations $\{ \bm{m}(x)  \}$ is approximated by a special set of configurations $\{ \bm{m}^{*}(x) = \xi^*(x) \mathcal{H} \xi^*(x)^\dagger\}$ so that 
\begin{align}
 W_C[U]   
\simeq \tilde{w}_C[U,\bm{m}^{*}]  ,
 \quad 
 \tilde\rho_C[U,\bm{m}^{*}] \simeq 1
  .
  \label{approx}
\end{align} 
For instance, this situation can occur for SU(2) by minimizing ${\rm tr} [(D_\mu[\mathscr{A}] \bm{m})^2]$ or ${\rm tr} [\mathbb{X}_\mu(x)^2 ]$ for a given configuration of $U_{x,\mu}$ (or $\mathscr{A}_\mu(x)$).

Now we are ready to consider the relationship between our approach and the conventional one based on the MAG for the issue of Abelian dominance. 
In the conventional MAG, after performing a local gauge transformation $\Omega_{x}$,   the link variable $U_{x,\mu}$   is decomposed to the diagonal link gauge field $u_\mu(x)$ and the remaining field $r_\mu(x)$:
\begin{equation}
 \tilde{U}_{x,\mu}:= \Omega_{x} U_{x,\mu} \Omega^{\dagger}_{x+\mu}
= r_\mu(x) u_\mu(x)
  ,
\end{equation}
where
\begin{equation}
 u_\mu(x) = {\rm diag.} (e^{i\theta_\mu^1(x)}, \cdots, e^{i\theta_\mu^N(x)}) ,
 \quad
 \sum_{a=1}^{N} \theta_\mu^a(x)= 0 \ (\text{mod} \ 2\pi)
  .
\end{equation}
The conventional Abelian(-projected) Wilson loop operator $W_C^{Abel}$ is defined by using the diagonal link gauge field $u_\mu(x)$ alone:
\begin{equation}
 W_C^{Abel}[u] = {\rm tr}\left\{ \prod_{<x,x+\mu>  \in C} u_\mu(x) \right\}/{\rm tr}(\bm{1}) 
 = \frac1N \sum_{a=1}^{N}  \exp \left\{ \sum_{<x,x+\mu> \in C} i\theta_\mu^a(x) \right\} 
  .
\end{equation}
In our formulation,  the diagonal part $u_\mu(x)$  is to be identified with $V_{x,\mu}$.  In fact,  $V_{x,\mu}$ is diagonalized by the transformation $ \xi^{*}_{x}{}^\dagger V_{x,\mu}\xi^{*}_{x+\epsilon \hat\mu}$, since $[\xi^{*}_{x}{}^\dagger V_{x,\mu}\xi^{*}_{x+\epsilon \hat\mu}, \mathcal{H}]=0$.
Therefore, we can identify  
\begin{equation}
u_\mu(x)= \xi^{*}_{x}{}^\dagger V_{x,\mu}\xi^{*}_{x+\epsilon \hat\mu} 
 .
\end{equation}
Indeed, the matrix element $\langle \xi^{*}_{x},\Lambda| V_{x,\mu} | \xi^{*}_{x+\epsilon \hat\mu}, \Lambda \rangle$ is a diagonal element of $\xi^{*}_{x}{}^\dagger V_{x,\mu}\xi^{*}_{x+\epsilon \hat\mu}$:
\begin{equation}
\langle \xi^{*}_{x},\Lambda| V_{x,\mu} | \xi^{*}_{x+\epsilon \hat\mu}, \Lambda \rangle
= \langle \Lambda| \xi^{*}_{x}{}^\dagger V_{x,\mu}\xi^{*}_{x+\epsilon \hat\mu} |   \Lambda \rangle 
= ( \xi^{*}_{x}{}^\dagger V_{x,\mu}\xi^{*}_{x+\epsilon \hat\mu})_{dd} 
  .
\end{equation}
 The diagonalization is achieved by rotating the color field $\bm{m}_{x}$ into a uniform direction on all sites:
\begin{equation}
 \xi^{*}_{x}{}^\dagger \bm{m}_{x} \xi^{*}_{x} 
 \rightarrow  \mathcal{H} := \bm{m}_0
    .
\end{equation}
This is why the degrees of freedom corresponding to the color field $\bm{m}_{x}$ do not exist in the conventional treatment of MAG. 
On the other hand, the remaining part $r_\mu(x)$  is to be identified with $X_{x,\mu}$. 
Combining these identifications with the above argument of an approximate Abelian dominance, we obtain the relation: 
\begin{equation}
\tilde{w}_C[U,\bm{m}^{*}=\bm{m}_0] \simeq W_C^{Abel}[u]  
  .
\end{equation}
In light of (\ref{approx}), 
this is interpreted as the approximate  Abelian dominance in the MAG observed in the conventional numerical simulation on a lattice. 
This result enables us to explain the reason why the infrared Abelian dominance in the string tension is not 100\% in the MAG on a lattice.  This is partly because there exists the lattice artifact coming from the non-zero lattice spacing, and partly because the specific configuration of the color field is chosen. 

\section{Conclusion and discussion}

In this paper, we have given a gauge-independent definition of  ``Abelian"  dominance in the  SU(N)  Wilson loop operator and  obtained a necessary and sufficient condition for achieving the exact ``Abelian" dominance
starting from the lattice regularization.  
In the continuum limit, we have shown that in the framework of decomposing the original SU(N) gauge field variable $\mathscr{A}_\mu(x)$ into two new variables $\mathscr{X}_\mu(x)$ and $\mathscr{V}_\mu(x)$:
\begin{equation}
 \mathscr{A}_\mu(x)=\mathscr{X}_\mu(x)+\mathscr{V}_\mu(x)
  ,
  \label{field-dec}
\end{equation} 
a set of conditions,  (\ref{condition-2}): ${\rm tr} \left\{ \mathscr{X}_\mu(x) \bm{m}(x) \right\}=0$ and  (\ref{DVm}): $D_\mu^{[\mathscr{V}]}  \bm{m}(x) = 0$, 
is a necessary and sufficient condition for the gauge-independent Abelian dominance for the Wilson loop operator in the fundamental representation defined as follows.
  
i) The Wilson loop operator $W_C[\mathscr{A}]$ defined by the line integral of the original gauge field $\mathscr{A}_\mu(x)$ is rewritten in terms of  the restricted (``diagonal") variable $\mathscr{V}_\mu(x)$ alone and there exist no contributions from the (``off-diagonal") variable $\mathscr{X}_\mu(x)$, i.e., 
\begin{equation}
W_C[\mathscr{A}]= {\rm const.} W_C[\mathscr{V}]
    .
\end{equation} 

ii) The line integral of $\mathscr{V}_\mu(x)$ along the loop $C$ is rewritten into the surface integral over the surface $S$ bounding the loop $C$ and the contribution to the surface integral comes from the component of the field strength $\mathscr{F}_{\mu\nu}[\mathscr{V}](x)$ (of the restricted variable $\mathscr{V}_\mu(x)$)  along the the ``Abelian" direction specified by $\bm{m}(x)$:  $F_{\mu\nu}:= 2{\rm tr}(\bm{m}(x)\mathscr{F}_{\mu\nu}[\mathscr{V}](x))$, i.e., 
\begin{equation}
W_C[\mathscr{V}]=W_S[F]
    .
\end{equation} 
The field decomposition (\ref{field-dec}) has been realized by way of a single color field $\bm{m}(x)$ for SU(N), whose explicit form  depends on the representation of the Wilson loop \cite{Kondo08}. 
For SU(2), in particular,  (\ref{condition-2}) and (\ref{DVm}) reduce to the first and second defining equations respectively for specifying the conventional CFNS decomposition \cite{Cho00}.

In the continuum limit, therefore, the ``Abelian"   dominance should be exact or 100\% realized, that is to say, the Wilson loop is exactly reproduced  from the restricted variable $\mathscr{V}_\mu(x)$ alone. 
This is the gauge-invariant statement of the ``Abelian" dominance.  
It should be remarked that the resulting exact ``Abelian" dominance in the continuum limit is an operator relation, which holds irrespective of the shape or the size of the loop $C$.  
Therefore the Abelian dominance shown in this paper is stronger than the conventional infrared ``Abelian" dominance.  
For instance, remember that the interquark potential $V(R)$ is obtained from the Wilson loop average $\left< W_C[\mathscr{A}] \right>_{\rm YM}$ in the Yang-Mills theory for a rectangular loop $C$ with side lengths $R$ and $T$, using the relationship:
$\left< W_C[\mathscr{A}] \right>_{\rm YM} \cong \exp [-TV(R)]$, or 
$V(R)=\lim_{T \rightarrow \infty} (-1/T) \ln \left< W_C[\mathscr{A}] \right>_{\rm YM}$.
In the conventional approaches, the ``Abelian" dominance was examined by measuring the potential, especially the string tension, i.e., the coefficient of the linear potential part in confinement phase where the linear potential part is dominant compared to a part of the Coulomb type in the long distance. This can be called the infrared ``Abelian" dominance in the confinement phase. 
The strong ``Abelian" dominance implies the equality of the Wilson loop itself holding irrespective of the phase in question.  Therefore, ``Abelian" dominance should hold also in the deconfinement phase in which there exists no linear potential.  
In this sense, ``Abelian" dominance does not guarantee quark confinement.

The decomposition (\ref{field-dec}) obeying (\ref{condition-2}) and  (\ref{DVm}) is able to extract the SU(N) gauge-invariant ``Abelian" part from the Yang-Mills theory without any approximations, which give a quite useful tool and a firm basis for discussing quark confinement based on the dual superconductor picture, since the dual superconductivity may be regarded as the electric-magnetic dual of the ordinary superconductivity described by an Abelian gauge theory. 
The monopole dominance should be considered separately based on a relationship between the Wilson loop and magnetic monopole, see e.g., \cite{Kondo08,KT99,KondoIV}.

Here we discuss a relationship between the strong ``Abelian" dominance and the nMAG constraint which is necessary to obtain the color field from the original gauge field. 
As a prescription for creating the color field $\bm{m}_{x}$ from the original gauge field configurations $U_{x,\mu}$, 
we have adopted the nMAG prescription. 
The nMAG is imposed by maximizing the functional
\begin{align} 
 F_{nMAG} 
 =& \sum_{x,\mu} \left\{ 1-
 Re{\rm tr} \left[  X_{x,\mu}  \right]/{\rm tr}(\bm{1}) 
 \right\} 
\nonumber\\
 =& \frac12 g^2 \epsilon^2 \sum_{x,\mu}  {\rm tr} [\mathbb{X}_\mu(x)^2 ]/{\rm tr}(\bm{1}) +  O(\epsilon^3) 
  .
\end{align}
We consider the Abelian dominance in the Wilson loop located at arbitrary position with arbitrary shape and size.
If we impose the nMAG, therefore, (\ref{prod-xi-xi}) is expected to have the minimum value.  
Under the nMAG, the deviation from the exact Abelian dominance is quite small to be $O(\epsilon^2)$ on a lattice.  
In other words, even on a lattice, the gauge-independent Abelian dominance holds to the accuracy $O(\epsilon^2)$.
By remembering that the nMAG reduces to the MAG under the gauge fixing $\bm{m}(x) \rightarrow \bm{m}_0$, i.e., the limit of uniform color field, 
this result explains why the Abelian dominance in the Wilson loop average was quite clearly observed even on a lattice under the conventional MAG. 
Thus we suggest that the nMAG plays the distinguished role as the best constraint for realizing the Abelian dominance in the Wilson loop average.

For SU($N$), $N \ge 3$, the above argument suggests the existence of a modified nMAG constraint obtained by minimizing the functional:
\begin{align} 
 \tilde{F}_{nMAG} 
 =& \sum_{x,\mu} \left\{ 1-
 Re{\rm tr} \left[  X_{x,\mu}  \right]/{\rm tr}(\bm{1}) 
-   2Re{\rm tr} \left[  X_{x,\mu}  \bm{m}_{x}   \right]
 \right\} 
\nonumber\\
 =& \frac12 g^2 \epsilon^2 \sum_{x,\mu} \left\{   {\rm tr} [\mathbb{X}_\mu(x)^2 ]/{\rm tr}(\bm{1}) 
+     {\rm tr} [\mathbb{X}_\mu(x)^2 \bm{m}(x)] \right\} +  O(\epsilon^3) 
 ,
\end{align}
which will enable us to obtain a better Abelian dominance than the previous one based on the nMAG. 
This issue will be discussed in more detail in a future publication.

Finally, we mention that there exist other approaches to examine the Abelian dominance based on miscellaneous gauges other than the maximal Abelian gauge, such as center gauge, Laplacian gauge, etc., see e.g., \cite{Greensite03}. 
Quite recently,  new approaches have also been proposed to achieve the gauge-independent Abelian dominance beyond the naive Abelian dominance mentioned in Introduction, see e.g., \cite{Suzuki}.  
It will be interesting to examine the relationship between these approaches and ours based on the viewpoint given in this paper.




\section*{Acknowledgments}
The authors would like to thank Takeharu Murakami and Toru Shinohara  for helpful discussions. 
This work is financially supported by Grant-in-Aid for Scientific Research (C) 18540251  from Japan Society for the Promotion of Science (JSPS).


\end{document}